\documentclass[prc,amsmath,twocolumn,superscriptaddress]{revtex4} 

\usepackage{dcolumn}
\usepackage{graphicx}
\usepackage{color}

\def\beq{\begin{equation}}
\def\eeq{\end{equation}}
\def\bea{\begin{eqnarray}}
\def\eea{\end{eqnarray}}

\renewcommand*{\eqref}[1]{Eq.~(\ref{eq:#1})}
\newcommand*{\eqlab}[1]{\label{eq:#1}}
\newcommand*{\figref}[1]{Fig.~\ref{fig:#1}}
\newcommand*{\figlab}[1]{\label{fig:#1}}

\newcommand*{\secref}[1]{Section~\ref{sec:#1}}
\newcommand*{\seclab}[1]{\label{sec:#1}}

\def\VYP#1#2#3{{\bf #1}, #3 (#2)}  

\def\NPB#1#2#3{Nucl.~Phys.~B~\VYP{#1}{#2}{#3}}

\def\PLB#1#2#3{Phys.~Lett.~B~\VYP{#1}{#2}{#3}}

\def\PRD#1#2#3{Phys.~Rev.~D~\VYP{#1}{#2}{#3}}
\def\PRL#1#2#3{Phys.~Rev.~Lett.~\VYP{#1}{#2}{#3}}

\newcommand{\etal}{\mbox{\textit et al.}}

\newcommand{\Omit}[1]{}

\begin{document}

\title{Determining neutrino absorption spectra at Ultra-High Energies.
}

\author{O. Scholten}
\email{scholten@kvi.nl}

\author{A. R. van Vliet}
\affiliation{Kernfysisch Versneller Instituut, University of Groningen, 9747 AA,
Groningen, The Netherlands}

\begin{abstract}
A very efficient method to measure the flux of Ultra-high energy (UHE) neutrinos
is through the detection of radio waves which are emitted by the particle shower
in the lunar regolith. The highest acceptance is reached for radio waves in the
frequency band of 100-200 MHz which can be measured with modern radio telescopes.
In this work we investigate the sensitivity of this detection method to
structures in the UHE neutrino spectrum caused by their absorption on the
low-energy relic anti-neutrino background through the Z-boson resonance. The
position of the absorption peak is sensitive to the neutrino mass and the
redshift of the source. A new generation of low-frequency digital radio
telescopes will provide excellent detection capabilities for measuring these
radio pulses, thus making our consideration here very timely.
\end{abstract}
\maketitle

\section{Introduction}

With the advent of a new generation of digital radio telescopes working in the
frequency range of 100-200 MHz it will become possible to measure the flux of
Ultra-High Energy (UHE) neutrinos, energies in excess of $10^{20}$~eV, as they
impact the Moon~\cite{Sch06}. To determine this flux as function of their energy
is of considerable interest as it may give some clues about their origin. The
recent results of the Pierre Auger Observatory on the correlation of UHE cosmic
rays with the positions in the sky of AGN~\cite{PAO07} have only increased the
interest.

UHE neutrinos will be absorbed on the low energy Cosmic Neutrino Background
(C$\nu$B) through the excitation of the Z-boson resonance~\cite{Weiler}. In
general this annihilation process is expected to lead to sizeable absorption dips
in the UHE neutrino spectra~\cite{Ringwald,Schrempp}. The positions of these
absorption lines depend on the redshift of the source emitting the UHE neutrinos.
The issue addressed here is the precision with which these absorption dips can be
determined in a realistic experiment. For the UHE background we will assume the
flux as given by the Waxman-Bahcall (WB) limit~\cite{Wax98}.

An exciting possibility is that the neutrino mass is not constant but varies with
redshift, Mass Varying Neutrinos (MaVaN). This has been proposed~\cite{Fardon} as
a model for dark energy. Such a neutrino mass variation introduces a distinctly
different structure of the absorption lines due to resonant absorption on relic
anti-neutrinos into Z-bosons than is obtained for redshift independent
masses~\cite{Ringwald,Schrempp}. In this work we will therefore also investigate
if this difference is observable in a realistic experiment. A very promising
method~\cite{Sch06} for measuring the flux of UHE neutrinos is through the
detection of the radio pulse which is emitted when UHE neutrinos initiate a
shower in the crust of the Moon.

In \secref{NuMoon} a short quantitative discussion of radio emission from
neutrinos hitting the lunar surface is presented. Following the approach of
Ref.~\cite{Sch06} we will calculate the pulse-height spectrum, as may be measured
at Earth. First we investigate in \secref{S-Energy} the detector response for
mono-energetic neutrinos. In \secref{E-Spectrum} the response to a spectrum of
neutrino energies is given using different scenarios for the neutrino mass and
redshift of the source. Using the statistics of a possible observation we show to
what extent the spectra for different neutrino-absorption scenarios may be
distinguished using a new generation of low-frequency digital radio telescopes
such as LOFAR~\cite{LOFAR} and SKA~\cite{SKA}. The conclusions are presented in
\secref{conclusions}.

\section{Radio emission}\seclab{NuMoon}

When an UHE particle induces a shower in dense media, such as ice, salt, and
lunar regolith, the front end of the shower has a surplus of electrons. Since
this cloud of negative charge is moving with a velocity which exceeds the
velocity of light in the medium, \v{C}erenkov radiation is emitted. For a
wavelength of the same order of magnitude as the typical size of this cloud,
which is in the radio-frequency range, coherence builds up and the intensity of
the emitted radiation reaches a maximum. This process is known as the Askaryan
effect~\cite{Ask62}. Dagkesamanskii and Zheleznyk~\cite{Dag89} were the first to
apply this mechanism to UHE neutrinos hitting the moon and proposed to use radio
telescopes at Earth to detect the emitted radiation which is coherent for radio
waves. Several experiments have since been performed~\cite{Han96,Gor04} to find
evidence for UHE neutrinos. All of these experiments have looked for this
coherent radiation near the frequency where the intensity of the emitted radio
waves is expected to reach its maximum. Since the typical lateral size of a
shower is of the order of 10~cm the peak frequency is of the order of 3~GHz.
Recently it has been proposed~\cite{Sch06} to perform such observations at an
even larger wavelength of about 3~m. This corresponds to the longitudinal extend
of the shower in the lunar regolith and as a result the angular spread of the
emitted signal is close to maximal while keeping coherence. As a result, the
detection efficiency for UHE particles is many orders of magnitude higher at
100~MHz than at 3~GHz.

The intensity of radio emission (expressed in units of Jansky's where
1~Jy~=~$10^{-26}$~W~m$^{-2}$Hz$^{-1}$) from a hadronic shower, with energy $E_s$,
in the lunar regolith, in a bandwidth $\Delta\nu$ at a frequency $\nu$ and an
angle $\theta$, can be parameterized as~\cite{Sch06}
\bea
&&\hspace*{-1.5em} F(\theta,\nu,E_s)
 = 3.86 \times 10^4\; e^{-Z^2} \Big( {\sin{\theta}\over \sin{\theta_c}} \Big)^2
 \Big( {E_s \over 10^{20} \mbox{ eV} } \Big) ^2
 \nonumber \\ &&\hspace*{-1.5em}  \times
 \Big( {d_{moon} \over d } \Big)^2
 \Big( {\nu \over \nu_0 (1+(\nu/\nu_0)^{1.44})} \Big)^2
 \Big({\Delta\nu \over 100\mbox{ MHz}} \Big) \; \mbox{Jy} \;,
 \eqlab{shower}
\eea
with
\beq
Z
 = (\cos{\theta} -1/n)
 \Big({n\over \sqrt{n^2-1}}\Big)\Big({180\over \pi \Delta_c}\Big)\;,
 \eqlab{Z}
\eeq
where $\nu_0=2.5$~GHz~\cite{Gor04}, $d$ is the distance to the observer, and
$d_{moon}=3.844 \times 10^8$~m is the average Earth-Moon distance. The angle at
which the intensity of the radiation reaches a maximum, the \v{C}erenkov angle,
is related to the index of refraction ($n$) of the medium, $\cos{\theta_c}=1/n$.
The parametrization of the energy dependence of the spreading of the radiated
intensity around the \v{C}erenkov angle, $\Delta_c$, is given in
Ref.~\cite{Sch06}.

The mean value for the attenuation length of radio waves in the regolith is taken
as $\lambda_r= (9/\nu$[GHz])~m. In the calculations we have included radiation
coming from a depth of at most 500~m treating for simplicity, and without loss of
accuracy~\cite{Sch06}, the whole layer as behaving like regolith. For the
neutrino an energy-dependent mean free path~\cite{Gan00} has been used,
$\lambda_\nu= 130\; \Big( {10^{20} \mbox{ eV} \over E_\nu } \Big)^{1/3}$~km,
which is appropriate for regolith.

For neutrino-induced showers only 20\% of the initial energy is converted to a
hadronic shower, while the remaining 80\% is carried off by the lepton. This
energetic lepton will not induce a detectable radio shower. For a muon, the
density of charged particles will be too small, while the shower of an UHE
electron will be elongated due to the Landau-Pomeranchuk-Migdal (LPM)
effect~\cite{Alv98}. The width of the \v{C}erenkov cone will thus be very small
which makes the shower practically undetectable. For the present calculations we
therefore have limited ourselves to the hadronic part of the shower. It should be
realized that the photons in the hadronic part of the shower have an energy that
is roughly 2 orders of magnitude less than the primary energy and thus that the
LPM effect is considerably smaller for this part.

\begin{figure}
    \includegraphics[height=7.9cm,bb=27 137 515 672,clip]{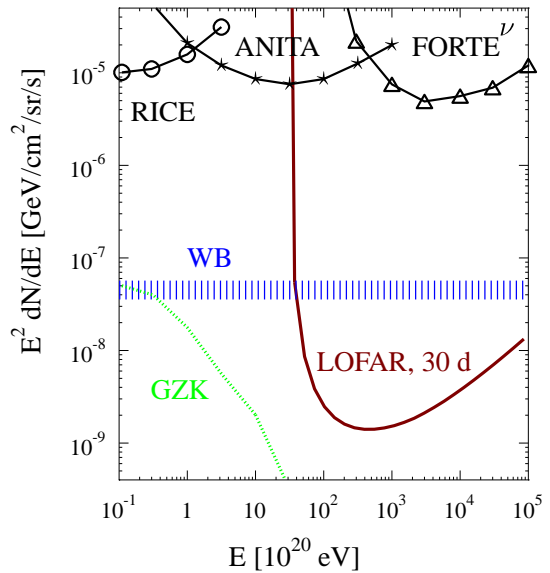}
\caption[fig6]{Flux limits, summed over all flavors, on UHE neutrinos as can be determined
with LOFAR observations (see text) are compared with
various models, in particular, WB~\cite{Wax98} (vertical bars),
GZK~\cite{Eng01} (dotted thin green line). Limits from the RICE~\cite{Kra03},
ANITA~\cite{Bar06}, and FORTE~\cite{Leh04} experiments are also shown.}
  \figlab{LOFAR}
\end{figure}

The radio flashes from the moon can be measured with powerful synthesis
telescopes such as LOFAR~\cite{LOFAR}. LOFAR can operate in the frequency band of
115-240~MHz where it will have a sensitivity of about $F_{\mbox{\tiny
noise}}=$20~Jy. Simulations show that a pulse of intensity 25$\times
F_{\mbox{\tiny noise}}$, interfering with the noise background, can be detected
with $3 \sigma$ significance at a probability greater than 80\%. For this reason
we have assumed in the calculations a detection threshold of 25$\times
F_{\mbox{\tiny noise}}$. A simulation for LOFAR, taking $\nu=165$~MHz, a
bandwidth of $\Delta\nu=100$~MHz, and an observation time of 30 days is shown in
\figref{LOFAR} assuming a 60\% Moon coverage. The result is compared with the WB
limit~\cite{Wax98}. With the future SKA telescope a sensitivity can be
reached~\cite{SKA} which is higher by several orders of magnitude.

The following analysis of the spectral sensitivity of the method is in principle
independent of the threshold sensitivity for detecting a radio signal. However
with a higher sensitivity a larger fraction of the pulse-height spectrum can be
determined with the possibility of increasing the spectrum-resolving power.

\section{Spectral resolving power}\seclab{spectrum}

An appealing opportunity to catch a glimpse of the C$\nu$B is through the energy
spectrum of UHE neutrinos. Such UHE neutrinos can annihilate with relic
anti-neutrinos (and vice versa) into Z bosons, if their energies coincide with
the respective resonance energies $E^{res}_{0,i}$ of the corresponding process
$\nu + \bar{\nu} \to Z$. These energies,
\beq
 E^{res}_{0,i} = {M^2_Z \over 2m_{\nu_{0,i}}} =
 4.2 \times 10^{21} {eV \over m_{\nu_{0,i}}} eV  \eqlab{Eres}
 \eeq
in the rest system of the C$\nu$B, are entirely determined by the Z boson mass
$M_Z$ as well as the respective neutrino masses $m_{\nu_{0,i}}$. An exceptional
loss of transparency of the UHE neutrinos results from the fact that the
corresponding annihilation cross-section on resonance is enhanced by several
orders of magnitude with respect to non-resonant scattering. As a consequence,
the diffuse UHE neutrino flux arriving at Earth should show absorption lines due
to this resonance scattering off the C$\nu$B, as explained in detail in Ref.
\cite{Weiler}. Our aim here is to determine the precision with which it is
possible to measure such absorption dips.  As reference or source spectrum, on
top of which the absorption is imposed, the WB limit~\cite{Wax98} is taken,
\begin{equation} \eqlab{WBlimit}
\Phi_{WB} = E^{2}\frac{dN}{dE}=
 3.5 \times 10^{-8}\mbox{\ [GeV cm$^{-2}$s$^{-1}$sr$^{-1}$]}\ .
\end{equation}
It is not a straightforward issue to determine the original neutrino energy
spectrum from the measured signals at Earth. When a neutrino interacts in the
lunar regolith, the magnitude of the generated electric pulse, as it arrives at
Earth, is only indirectly related to the original energy, as can be seen from
\eqref{shower}. The strength of the pulse depends strongly on emission angle and
the energy fraction of the hadronic part of the shower~\cite{Bev07}. In addition
the signal is attenuated in the lunar regolith. Although it may be possible to
determine emission angle and depth from the frequency dependence of the pulse, we
do not want to rely on this in the present investigation. To calculate the
sensitivity of the NuMoon measurements to the energy spectrum of UHE neutrinos,
as a first step the pulse-height spectra are calculated as received at Earth from
mono-energetic neutrinos, integrated over angles of incidence, emission angles,
and emission depth. In the second step these pulse-height spectra are integrated
over the neutrino-energy distributions.

\subsection{Single energy}\seclab{S-Energy}

\begin{figure}[tb]
\centerline{
\scalebox{0.35}{\includegraphics{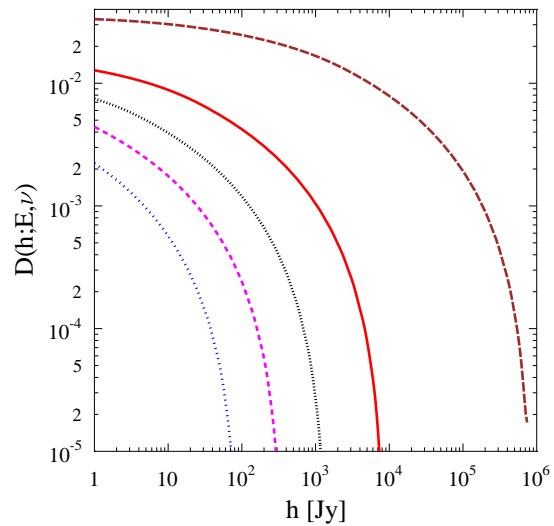}}}
\caption{Detection probability ${\cal D}(h;E)$ v.s.\ threshold $h$
for different neutrino energies $E$. Blue, dotted, line: $E=10^{21}$~eV,
magenta, dashed, line: $E=2 \times 10^{21}$~eV,
black, close-dotted, line: $E=4 \times 10^{21}$~eV,
red, drawn, line: $E=10^{22}$~eV, brown, close-dashed, line: $E=10^{23}$~eV.}
\figlab{detthrvslum}
\end{figure}

To be able to address quantitatively the pulse-height spectrum which is
measurable at Earth, we introduce the detection probability ${\cal D}(h;E)$. It
is defined as the probability that a neutrino hitting the Moon at an arbitrary
angle and position with energy $E$ will produce radio waves at a frequency of
$\nu=100$~MHz, with a power at Earth exceeding a certain threshold value $h$ in a
bandwidth of 50~MHz. In \figref{detthrvslum} detection probabilities are given as
a function of the detection threshold $h$ for different neutrino energies in the
range of interest for this paper.

\begin{figure}[tb]
\scalebox{0.35}{\includegraphics{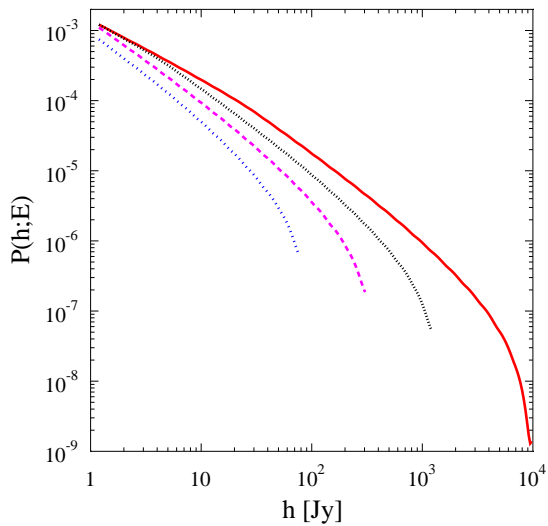}}
\caption{Probability distribution $P(h;E)$ for neutrinos with the same energies
as used in \figref{detthrvslum}.}
\figlab{pulsehvsprob}
\end{figure}

The probability ${\cal P}(h;E)$ for measuring a signal with a pulse height
between $h-0.5$ and $h+0.5$ when a neutrino with energy E impinges on the surface
of the Moon, can now be obtained by differentiating ${\cal D}(h;E)$ with respect
to the detection threshold $h$. The obtained functional dependence of ${\cal
P}(h;E)$ on $h$ is shown in \figref{pulsehvsprob}.

\begin{figure}[tb]
\scalebox{0.35}{\includegraphics{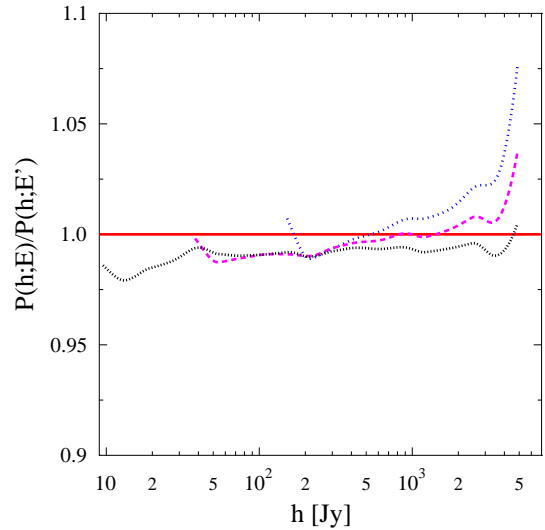}}
\caption{Ratio of the exact probabilities at E=$10^{22}$~eV with those reconstructed
by applying Eq.~\ref{eq:genprobvsA},
from the results at lower energies; blue, dotted, line from E'=$10^{21}$~eV,
magenta, dashed, line from E'=$2 \times 10^{21}$~eV and
black, dashed-dotted, line from E'=$4 \times 10^{21}$~eV. }
\label{fig:endepgen}
\end{figure}

The maximum pulse height which can be detected at Earth emitted by a moon shower
which is induced by a neutrino with an energy $E$ is written as $h=A(E)$. This
maximum follows from the strength of the emitted radio pulse as given by
\eqref{shower} \Omit{gives 152 Jy at 120 MHz and E=$10^{21}$} plus some 20\%
attenuation at the lunar surface,
\begin{equation} \eqlab{asymptote}
A(E)=128\ \mbox{Jy} \times \left(\frac{E}{10^{21}\ \mbox{eV} }\right)^2  ,
\end{equation}
and this reproduces the end-point values given in \figref{detthrvslum}. Of
particular interest here is the energy dependence which will be useful to
parameterize the energy dependence of $P(h;E)$. Numerical results indicate that
$P(h;E)$ scales with energy as
\begin{equation} \label{eq:genprobvsA}
P(h;E) 
 = P(h';E') \times \left(\frac{E'}{E}\right)^{1.73}
\end{equation}
with
\beq
h'={A(E') \over A(E) } h = \left( {E' \over E}\right)^2 h \;, \eqlab{hvshp}
\eeq
where $A(E)$ is given by \eqref{asymptote}.  To show the accuracy of this
parametrization we plot in \figref{endepgen} the ratio of the exact probabilities
at E=$10^{22}$~eV with those reconstructed from the results at lower energies.
Considering the simplicity of the parametrization over several orders of
magnitude, the agreement is excellent.

\subsection{The neutrino absorption spectrum} \seclab{E-Spectrum}

The absorption of UHE neutrinos in the intergalactic space is generally small and
the neutrinos arriving at Earth may have originated from sources at great
distances. In the calculation of the absorption dips in the spectrum it is
therefore essential to include in the consideration the expansion of the universe
and thus the redshift of far-away sources. In the following we discuss the
structure of the absorption dips in the neutrino spectrum in two different
pictures, first for the picture in which the neutrino mass is independent of
redshift followed by that for the MaVaN picture~\cite{Fardon}.

\begin{figure}[htb]
\centerline{ \scalebox{0.35}{\includegraphics{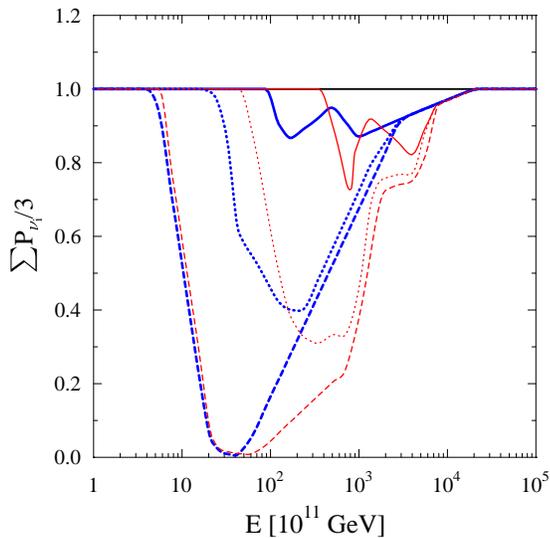}} }
\caption{The averaged neutrino survival probability, $1/3 \sum P_{\nu_{i}}$,
as a function of neutrino energy (extracted from Fig.~5 of Ref.~\cite{Ringwald})
for sources at different
redshifts; $z_s=5$ (drawn line), $z_s=20$ (dotted line), $z_s=50$ (dashed line).
The thick, blue, curves give the absorption spectra for the case that the neutrino
mass is independent of redshift, while the thin, red, curves gives the predictions
in the MaVaN model.
\figlab{spectrum} }
\end{figure}

For a redshift independent neutrino mass the width of the absorption peaks
depends strongly on the redshift of the source, $z_{s}$. The peak position is
determined by the neutrino mass. Without considering thermal broadening, which is
a small effect, resonant absorption occurs for the neutrinos in the energy
interval~\cite{Ringwald}
\bea
\frac{E^{res}_{0,i}}{1+z_s}<E_{0}<E^{res}_{0,i} \eqlab{interval}
\eea
where $ E^{res}_{0,i}$ is given in \eqref{Eres}. Absorption at energies below
$E^{res}_{0,i}$ occurs at large distances where the neutrino energy is larger
than observed at Earth due to the redshift. The thick, blue, curves in
\figref{spectrum} give the neutrino absorption spectra, using the values given in
Fig.~5 of Ref.~\cite{Ringwald}, in the picture that the neutrino masses are
independent of redshift, the static-mass neutrino (StaMaN) picture. The results
are given for three different distances of the source, corresponding to redshifts
of $z_{s}=5$, $z_{s}=20$ and $z_{s}=50$. A normal hierarchy for neutrino masses
has been used, $m_{\nu 1}=10^{-5}$~eV, $m_{\nu 2}=8.3 \times 10^{-3}$~eV and
$m_{\nu 3}=5.17 \times 10^{-2}$~eV. The absorbtion lines are calculated for a
universe with a day matter density $\Omega_M = 0.3$, a curvature density
$\Omega_k = 0$ and vacuum energy density $\Omega_\Lambda = 0.7$ as corresponds to
a Lambda Cold Dark Matter ($\Lambda$CDM) universe as suggested by recent
observations~\cite{Kom08} with a homogeneous relic neutrino density.

The absorption lines are superimposed on a smooth neutrino spectrum with a flux
equal to the WB-limit. From \figref{spectrum} one sees clearly that for larger
values of $z_s$ the absorption dips show at smaller neutrino energies, as
predicted based on \eqref{interval}. In addition the absorption is getting more
pronounced because of the larger travelling distance of the UHE neutrinos through
the C$\nu$B. Structures in \figref{spectrum} at lower energies shown are due to
the absorption of neutrinos with higher mass. Due to neutrino oscillations the
absorption at lower energies can be almost complete, instead of only 1/3 when
only neutrinos of a single flavor are absorbed. Most important for the present
study are the heavier mass neutrinos which generate structures in the spectrum at
lower energies where the sensitivity for detecting it is largest due to the
higher flux.

In the MaVaN model the absorption dips in the neutrino spectrum become
independent of $z_{s}$ for small $z_{s}$ because $m_{\nu i}(z) \propto
(1+z)^{-1}$. This implies that for relatively nearby sources of UHE neutrinos one
expects to see sharp absorption lines in the spectrum, given by the thin, red,
curves in \figref{spectrum}. For sources at distances corresponding to $z_s >10$
the predicted redshift dependence is $m_{\nu i}(z) \propto (1+z)^{-1/2}$ and as a
result the lines broaden considerably. At a redshift of $z_s=50$ this effect is
so big that the differences in the absorption spectra in the MaVaN picture and
the StaMaN picture are minor~\cite{Ringwald} at smaller neutrino energies.

\begin{figure}[tb]
\centerline{
\scalebox{0.35}{\includegraphics{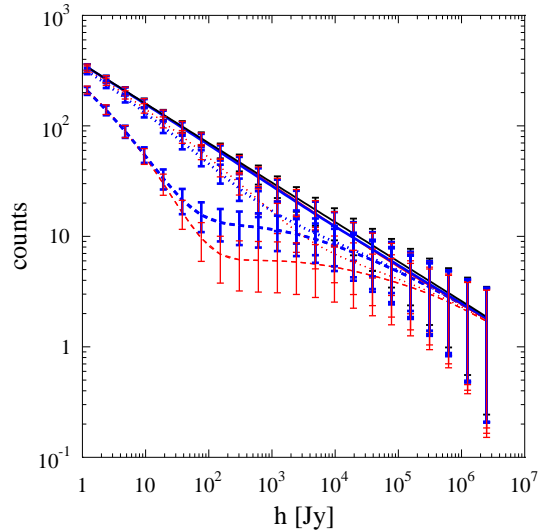}}}
\caption{The expected count rate for a realistic experiment for the different
neutrino
absorption dips given in \figref{spectrum}.}
\figlab{counts}
\end{figure}

Based on the detection probability for pulses at Earth, the neutrino flux, and
the length of a typical experiment, the expected number of radio pulses of a
particular magnitude can be calculated. For the present estimates given in
\figref{counts} the typical measuring conditions with LOFAR, as discussed in
Ref.~\cite{Sch06}, are assumed, i.e.\ a measuring time of 30 days where the
telescopes have only half the moon is in the field of view. The quoted error bars
reflect the statistical error only. Of course for longer observing times or
better coverage of the moon the statistics will improve. The bin size of the
pulse-height spectrum is chosen the same as was assumed in \figref{pulsehvsprob}
where the size of the bin is taken equal to the pulse height.

From \figref{counts} it can be seen that in a realistic experiment one is
sensitive to the absorption structures in the spectra only if the sources are at
the largest redshifts. Even though the absorption structures in the neutrino
spectra are distinctly different, these structures are close to impossible to
unfold from the measured pulse-height spectrum. One thus concludes that, on the
basis of these data, it will not be possible to distinguish the MaVaN and the
StaMaN pictures. On the other hand the measured values are clearly sensitive to
the source distance, independent of the picture used for the neutrino mass.

\begin{figure}[tb]
\centerline{
\scalebox{0.35}{\includegraphics{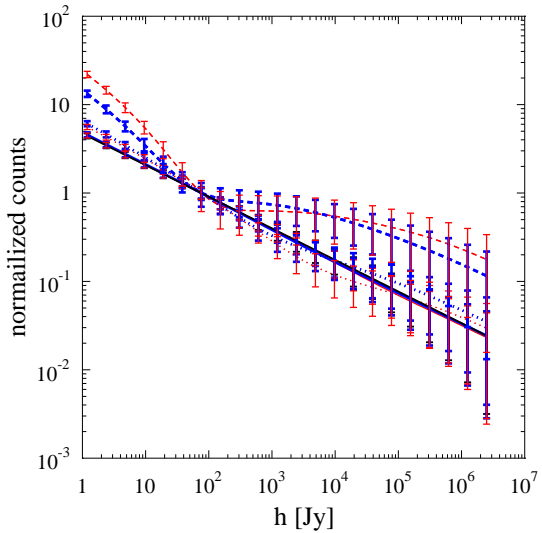}}}
\caption{The counts given in \figref{counts} are re-normalized to those at
a pulse height of $80~Jy$ to the value expected from the spectrum without absorption.}
\figlab{dNdHnorm}
\end{figure}

For a realistic observation, however, the absolute neutrino flux in not known and
the absolute number of counts can not be used to distinguish the different
features in the neutrino spectra. For this reason we compare the spectra where
the counts have been normalized to the same value at a relatively low pulse
height, $h=4\times F_{\mbox{\tiny noise}}=80$~Jy in the present example, which
should be typical for a LOFAR observation. The normalized spectra are shown in
\figref{dNdHnorm}. From this figure it can be seen that the observation will not
be able to convincingly distinguish between the MaVaN and the StaMaN models. One
should however be able to determine the absorption lines if the sources are at
distances of the order of $z_s=50$. Seeing the absorption lines would confirm
that indeed a C$\nu$B exists. In observations with SKA, where the higher
sensitivity allows to distinguish even smaller pulses from the background, one
may be able to see absorption structures with sources as close as $z_s=20$.

\section{Summary}\seclab{conclusions}

Due to resonant absorption on the C$\nu$B the energy spectrum of UHE neutrinos
carries with it the information on the neutrino mass and the distance of the
emitting source from the Earth. As stressed already in Ref.~\cite{Ringwald} the
structure of the absorption lines is markedly different in a model where the
neutrino mass is constant as compared to that predicted in the MaVaN
model~\cite{Fardon}. In this work we have addressed the question whether these
differences can be determined in a realistic experiment. The most realistic
possibility to observe the small flux expected for UHE neutrinos is by observing
the radio waves in the frequency range of 100-200~MHz that are emitted when these
neutrinos impact the Moon~\cite{Sch06}. Since the observed pulse height is only
indirectly related to the primary energy of the neutrino we have made a realistic
simulation of such a measurement based on the neutrino absorption calculated in
Ref.~\cite{Ringwald}. The results show that, due to the smallness of the
absorption or, equivalently, due to the high degree of transparency of
intergalactic space for UHE neutrinos, absorption lines are hardly visible for
sources closer than $z_s=20$. For sources at a redshift of $z_s=50$, the effects
of the absorption on the C$\nu$B is clearly visible in the measurement. The
predicted differences between the two models for the neutrino mass are too small
to be detectable in a convincing way.

\begin{acknowledgments}
This work was performed as part of the research programs of the Stichting voor
Fundamenteel Onderzoek der Materie (FOM), with financial support from the
Nederlandse Organisatie voor Wetenschappelijk Onderzoek (NWO).
\end{acknowledgments}

\end{document}